\documentclass{article}

\usepackage[english]{babel}

\usepackage[letterpaper,top=2cm,bottom=2cm,left=3cm,right=3cm,marginparwidth=1.75cm]{geometry}

\usepackage{amsmath}
\usepackage{graphicx}
\usepackage[colorlinks=true, allcolors=blue]{hyperref}

\usepackage{xcolor}
\usepackage{comment}
\usepackage{hyperref}
\usepackage{caption}
\usepackage{amsmath}
\usepackage{bbold}
\usepackage{etoolbox}
\usepackage{adjustbox}
\usepackage{makecell}
\usepackage{subcaption}
\usepackage{textcomp}
\usepackage[symbol]{footmisc}
\usepackage{booktabs}
\usepackage{afterpage}
\usepackage{algorithm}
\usepackage{algorithmic}
\usepackage{setspace}
\usepackage{boldline}
\usepackage{bbm}
\usepackage{multirow}
\usepackage[version=4]{mhchem} 
\usepackage{array}   
\newcolumntype{L}{>{$}c<{$}} 
\usepackage[sort&compress,numbers]{natbib}
\providecommand{\keywords}[1]{\textbf{Keywords:} #1}

\title{Investigating the Surrogate Modeling Capabilities of Continuous Time Echo State Networks}

\author{\stepcounter{footnote}Saakaar Bhatnagar\thanks{Corresponding Author} \\ {\small \textit{saakaar@stanford.edu}} }

\begin{document}

\maketitle

 \\

\begin{abstract}
  \centering Continuous Time Echo State Networks (CTESNs) are a promising yet under-explored surrogate modeling technique for dynamical systems, particularly those governed by stiff Ordinary Differential Equations (ODEs). A key determinant of the generalization accuracy of a CTESN surrogate is the method of projecting the reservoir state to the output. This paper shows that of the two common projection methods (linear and nonlinear), the surrogates developed via the nonlinear projection consistently outperform those developed via the linear method. CTESN surrogates are developed for several challenging benchmark cases governed by stiff ODEs, and for each case, the performance of the linear and nonlinear projections is compared. The results of this paper demonstrate the applicability of CTESNs to a variety of problems while serving as a reference for important algorithmic and hyper-parameter choices for CTESNs. 
\end{abstract}

\section{Introduction}

\label{sec:intro}

 Modeling dynamic systems using scientific machine learning (SciML) techniques is a rapidly growing field with advanced ML architectures being applied to model complex problems across a diverse range of applications. Some examples are rapid design optimization \cite{nascimento2023framework}, real-time health monitoring \cite{malekloo2022machine}, turbulent flow modeling \cite{yousif2023deep}, and materials discovery \cite{liu2017materials}. Many of these applications utilize a " surrogate model " that makes real-time predictions of the system behavior in place of full-order models that would be too slow or expensive for the application.

Recently, Echo State Networks (ESNs) have seen an increase in popularity for modeling highly nonlinear and chaotic phenomena in domains such as optimal control planning \cite{liang2023online}, chaotic time series prediction \cite{li2022echo,xu2018hybrid}, signals analysis \cite{duarte2024denoising} and even turbulent fluid flow \cite{ghazijahani2022benefits}. These applications leverage the ability of ESNs to capture highly nonlinear transient behavior accurately, as well as the extremely low cost of training them, with the empirical success of ESNs on a wide range of approximation tasks discussed and explained in several works \cite{gonon2023approximation,hart2020embedding}. One of the biggest limitations, however, for using the standard ESN implementation in the surrogate modeling of nonlinear dynamical systems, is that the available training data may not be uniformly sampled in time. A particular example of this is the numerical solution of stiff systems of Ordinary Differential Equations (ODEs) from ODE solvers. An ODE system is said to be stiff if, for any initial conditions and in certain intervals, the solving numerical method is forced to use a timestep which is very small compared to the smoothness of the exact solution \cite{lambert1991numerical}. Numerical ODE solvers overcome the instability due to stiffness by having a variable time step size during the solve, leading to uneven sampling of the solution in time.

There have been several attempts to apply reduced-order modeling for stiff ODEs. \citet{ji2021stiff} used Physics Informed Neural Networks \cite{cuomo2022scientific}, \citet{kim2021stiff} used Neural ODEs \cite{chen2018neural} and \citet{goswami2024learning} used Deep Operator Networks\cite{lu2021learning} to solve several stiff systems such as the ROBER \cite{gobbert1996robertson}  and POLLU \cite{verwer1994gauss} problems. However, the methodologies applied require assumptions and scalings that may not generalize, or require lots of training data and compute resources to train deep neural networks. Work by \citet{anantharaman2020accelerating} also showed the failure of popular architectures such as Long Short Term Memory (LSTM) and standard ESNs in modeling stiff systems.

To use the attractive properties of ESNs (i.e capacity to model highly nonlinear signals and ease of training) for modeling stiff systems, a variant of ESNs called Continuous Time Echo State Networks or CTESNs has been proposed \cite{anantharaman2020accelerating} to address this issue. CTESNs have been successfully employed in various applications from accelerating predictions of power system dynamics \cite{roberts2022continuous} to accelerating solutions of pharmacology models \cite{anantharaman2022stably}.  
 
 In recent literature, two ways of using CTESNs for surrogate modeling have emerged; the Linear Projection CTESN (LPCTESN) \cite{anantharaman2020accelerating} and the Nonlinear Projection CTESN \\(NLPCTESN) \cite{roberts2022continuous,anantharaman2022stably} and have been applied to a range of problems. However, there is currently a lack of work critically examining the accuracy of both methods and how they compare to one another. Further, both projection methods use a Radial Basis Function (RBF) for interpolation, which also comes with several algorithmic choices that need to be explored. This study aims to fill this gap by investigating the effects of these algorithmic choices on surrogates created to solve several stiff ODE systems such as Robertson's equations, the Sliding Basepoint model of automobile collision, and the POLLU air pollution model. The findings of the study show that for the same hyper-parameter settings of the CTESN, the NLPCTESN outperforms the LPCTESN on all benchmarks shown. Further, it is shown that for the interpolating RBFs used within CTESNs, k-Nearest Neighbor (k-NN) polynomial augmented RBFs outperform standard RBFs in predictive accuracy.

 This paper is divided as follows: Section \ref{sec:ESN} introduces the concept of ESN and CTESN, along with the projection methods LPCTESN and NLPCTESN described above discussed in detail. Section \ref{sec:applications} demonstrates the application of the methods to several challenging stiff ODE problems, with a qualitative and quantitative discussion of the results. Section \ref{sec:conclusion} summarizes the work with possible future directions to take.

\section{Methods}
\label{sec:ESN}

\subsection{Standard Echo State Networks}

\begin{figure}[h!]
    \centering
    \includegraphics[angle=0,width=10 cm]{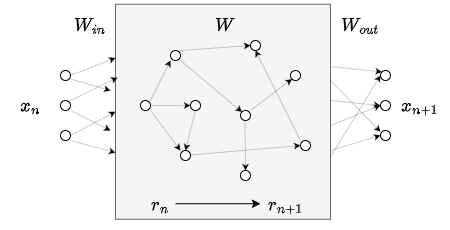}
    \caption{\centering Depiction of a standard Echo State Network.}
    \label{fig:ESN_diagram}
    \centering
    
\end{figure}

An Echo State Network \cite{jaeger2007echo}, depicted in Figure \ref{fig:ESN_diagram}, is a form of reservoir computing that is very similar to the more popular Recurrent Neural Network (RNN) in its architecture. They make predictions by following a recurrence relation (Equation \ref{eqn:ESN_main}) for updating a latent space vector and using that vector to map to a given output. Unlike RNNs, the parameters of the RNN in the " reservoir " are fixed and are not updated during training, and it is only the mapping from latent space to the output space that is learned. Depending on the implementation, this makes the training of ESNs very fast (sometimes as fast as a simple least squares fit) and computationally cheap.

The governing equation for an echo state network reads:

\begin{equation}
\label{eqn:ESN_main}
    \textbf{r}_{n+1}= \sigma (\textbf{W}_{in}\textbf{x}_{n}+\textbf{W} \textbf{r}_{n}), 
\end{equation}

where \(\textbf{r}_{n} \in \mathbb{R}^{N_{r}}\) is the latent state at timestep n, \(\sigma\) is an activation function (most commonly tanh), \(\textbf{W}_{in} \in \mathbb{R}^{N_{r}\times N_{x}}\) and \(\textbf{W} \in \mathbb{R}^{N_{r}\times N_{r}}\) are the reservoir matrices which are fixed and randomly initialized, and \(x_{n}\) is the system state at time n. The matrix \textbf{W} is a sparse matrix and usually has 1\% nonzero entries.

 The output projection reads:

\begin{equation}
\label{eqn:ESN_proj}
    \textbf{x}_{n+1}= \Phi(\textbf{r}_{n+1}),
\end{equation}

where \(\Phi\) is decided by the projection method. The most popular projection method is the linear projection, resulting in 

\begin{equation}
\label{eqn:ESN_linear_proj}
    \textbf{x}_{n+1}= \textbf{W}_{out}\textbf{r}_{n+1},
\end{equation}

where \(\textbf{W}_{out} \in \mathbb{R}^{N_{x}\times N_{r}}\) is the linear projection matrix that needs to be fitted to the training data. Like most machine learning algorithms, ESNs have a set of hyper-parameters that need to be tuned, and there are several works \cite{viehweg2023parameterizing,lukovsevivcius2012practical} that can be used as guides to select them. A key hyper-parameter in ESNs is the spectral radius of \(\textbf{W}\). In this study, the spectral radius is fixed to a value of (0.01) for all models created. Although a bit smaller than the conventional values used in standard ESNs, it was found in this study on CTESNs that using the slightly smaller value maximized predictive accuracy while also ensuring the stability of the reservoir (\textbf{r}) solution.
To fit the trainable matrix, one only has to solve the ordinary linear least squares problem:

\begin{equation}
\label{eqn:least_sq_fit}
    \begin{aligned}
    \begin{split}
    \text{If }X=[\textbf{x}_{1}; \textbf{x}_{2};....; \textbf{x}_{N}] &\text{ and } R=[\textbf{r}_{1}; \textbf{r}_{2};.....; \textbf{r}_{N}]\text{ then}\\
    \textbf{W}_{out}&= (RR^{T})^{-1}RX.
    \end{split}
    \end{aligned}
\end{equation}

\subsection{Continuous Time Echo State Networks (CTESN)}

Continuous Time Echo State Networks are a variant of ESNs that model time as a continuous rather than discrete quantity. The model equation for a CTESN is given by:

\begin{equation}
    \label{eqn:CTESN_main}\dot{\textbf{r}}=\sigma(\textbf{W}_{in}\textbf{x}+\textbf{W} \textbf{r}),
\end{equation}
with all variables having the same definition as in the previous section. The projection equation reads:

\begin{equation}
\label{eqn:NLPCTESN_proj}
     \textbf{x}= \Phi(\textbf{r}).
\end{equation}

As per the literature, there are two ways of modeling the relation between the latent space and the outputs \cite{rackauckas2022composing}; the linear method (called Linear Projection CTESN or LPCTESN), described by:

\begin{equation}
    \textbf{x}= \textbf{W}_{out}\textbf{r}.
\end{equation}

The second method is the nonlinear method (called NonLinear Projection CTESN or NLPCTESN), where the projection \(\Phi\) is a nonlinear mapping. Many possible functions can be used, but the literature on NLPCTESNs \cite{roberts2022continuous,rackauckas2022composing} currently uses standard Radial Basis Functions (RBFs) to write the projection:

\begin{equation}
    \textbf{x}= \textbf{RBF}(\textbf{r}).
\end{equation}

\subsubsection{Surrogate Modeling via CTESNs}

 To create and use a surrogate model via the CTESN method, a few steps are followed. First, a Design of Experiments (DoE) space is created and N sample combinations of query parameters are drawn; call this set P=\{\(p_{1},p_{2}.....p_{N}\)\}. Each \(p_{i}\) represents a set of conditions at which the ODE is solved and the surrogate is expected to capture the change in the solution due to changes in the value of \(p_{i}\). Examples include the initial conditions (see Sections \ref{sec:rober_init} and \ref{sec:sliding_basepoint}), rate constants of the ODE (see Sections \ref{sec:rober_rate} and \ref{sec:POLLU}), etc. The ODE is solved numerically at each \(p_{i} \in \) P, to return the solution set Y=\{\(\textbf{y}_{1},\textbf{y}_{2}....\textbf{y}_{n}\)\}, \(\textbf{y}_{i} \in \mathbb{R}^{N_{x}\times N_{ts}^{i}}\). A single parameter combination \(p^{*} \in \) P is drawn at random, and the reservoir ODE (given by Equation \ref{eqn:CTESN_main}) is solved using a numerical ODE solver. This returns a \textbf{r}(t) \(\in \mathbb{R}^{N_{r}\times N_{ts}}\) where \(N_{ts}\) is the number of timesteps in the solution at \(p^{*}\). Then, depending on whether the method follows a linear or nonlinear projection method, either Algorithm \ref{linear_proj_steps} or Algorithm \ref{nonlinear_proj_steps} is followed to fit and query the surrogate.

\begin{algorithm}
    \caption{Linear Projection CTESN Surrogate Fitting}\label{linear_proj_steps}
    \begin{algorithmic}
     \setstretch{1.35}   
        \FOR{\(\textbf{y}_{i}\) in Y}
            
            \STATE Fit \(W_{out}^{i}\) from \(\textbf{y}_{i}=W_{out}^{i}\cdot\) \textbf{r} using Equation \ref{eqn:least_sq_fit}
        \ENDFOR

        \STATE Fit RBF mapping \(W_{out}^{i}=\textbf{RBF}(p_{i}), \forall\) (\(W_{out}^{i},p_{i}\)) pairs

        \STATE To query a new parameter \(p_{test}\):

        \STATE \textbf{    Step 1}: \(W_{out}^{test}=\textbf{RBF}(p_{test})\)

        \STATE \textbf{    Step 2}: \(\textbf{y}_{test}\)(t)=\(W_{out}^{test} \cdot\)\textbf{r} 
   
    \end{algorithmic}
\end{algorithm}

\begin{algorithm}
    \caption{NonLinear Projection CTESN Surrogate Fitting}\label{nonlinear_proj_steps}
    \begin{algorithmic}
     \setstretch{1.35}   
        \FOR{\(\textbf{y}_{i}\) in Y}
            
            \STATE Fit \(\textbf{RBF}_{1}^{i}\) as per \(\textbf{y}_{i}=\textbf{RBF}_{1}^{i}(\textbf{r}(t),\textbf{W}_{RBF_{1}}^{i})\)
        \ENDFOR

        \STATE  Fit \(\textbf{RBF}_{2}\) as per \(\textbf{W}_{RBF_{1}}^{i}=\text{RBF}_{2}(p_{i})\), \( \forall\)  \((\textbf{W}_{RBF_{1}}^{i}, p_{i})\) pairs

        \STATE To query a new parameter \(p_{test}\):

        \STATE \textbf{    Step 1}: \(W_{RBF_{1}}^{test}=\textbf{RBF}_{2}(p_{test})\)

        \STATE \textbf{    Step 2}: \(\textbf{y}_{test}\)(t)=\( \textbf{RBF}_{1}(W_{RBF_{1}}^{test},\textbf{r}\)(t))
   
    \end{algorithmic}
\end{algorithm}

Algorithms \ref{linear_proj_steps} and \ref{nonlinear_proj_steps} both use an interpolating RBF in their procedure, and in this article, it is demonstrated that polynomial augmented k-nearest neighbor (k-NN) RBFs deliver superior results in terms of generalization to new test problems for CTESN-based surrogate models, compared to standard RBFs. The reader is encouraged to read Appendix Section \ref{sec:NN_RBF_Poly} for a detailed explanation of k-NN polynomial augmented RBFs.

\section{Applications and Results}
\label{sec:applications}

CTESN surrogates are created for three problems. First Robertson's equations are solved, parametrizing the rates of reaction and initial condition separately. Then, the Sliding Basepoint system is modeled, comparing explicitly the difference due to the projection method, and parametrizing the initial conditions of the problem. Finally, the POLLU system is solved, again parametrizing the initial conditions of the problem and comparing the results obtained using differing projection methods. Unless mentioned otherwise, the training data was sampled from the DoE space randomly.\\

In all results in this section, the MAE is computed as 

\begin{equation}
    \textbf{MAE} = \frac{1}{N_{test}}\sum_{j=1}^{N_{test}}\frac{\sum_{i=1}^{N_{ts}^{j}}|y_{j,pred}^{i}- y_{j,true}^{i}|}{N_{ts}^{j}},
\end{equation}

where \(N_{test}\) is the number of test cases, \(y_{j,pred}^{i}\) and \(y_{j,true}^{i}\) are the prediction and true solution respectively for the i'th time step of the j'th test case. 

\subsection{Robertson's Equation}

The first demonstrated application is the surrogate modeling of the Robertson Equations \cite{gobbert1996robertson}. They are written as:
\begin{equation}
\begin{aligned} 
    \begin{split}
    \dot{y_{1}}&=-r_{1}\cdot y_{1}+r_{3}\cdot y_{2}\cdot y_{3},\\
    \dot{y_{2}}&=r_{1}\cdot y_{1}-r_{3}\cdot y_{2}\cdot y_{3}-r_{2}\cdot y_{2}^{2},\\
    \dot{y_{3}}&=r_{2}\cdot y_{2}^{2},
    \end{split}
\end{aligned}
\end{equation}
where \(y_{1},y_{2}\) and \(y_{3}\) represent the concentration of the reactant species and \(r_{1},r_{2}\) and \(r_{3}\) represent the fixed rates of reaction. The system is usually subject to initial conditions of \([y_{1},y_{2},y_{3}]=[1,0,0]\).

They are a system of ODEs that describe a standard rate process and are often used to benchmark numerical ODE solvers due to the stiffness of the system. More specifically, the long-time integration of Robertson's equations is known to be a challenging problem for numerical ODE solvers, and the system hence serves as a good test for surrogate models. In this work models are created by parametrizing the system in two ways; first, the rates of reaction are parametrized. This has been the focus of several other papers written on CTESNs \cite{anantharaman2020accelerating,rackauckas2022composing}. Second, the initial conditions of the system are parametrized.

\subsubsection{Parametrizing Rates of Reaction}
\label{sec:rober_rate}

In this section, the Design of Experiment (DoE) space for the rates is given as:

\begin{equation}
\begin{aligned}
    \begin{split}
    r_{1}&=[0.032,0.048],\\
    r_{2}&=[2.4,3.6]\times E7,\\
    r_{3}&=[0.8,1.2]\times E4.\\
    \end{split}
\end{aligned}
\end{equation}

The focus is limited to presenting the results of the prediction of the variable \(y_{2}\). This variable has a sharp transient that occurs at a time scale much smaller than the other two, and hence causes the system to be stiff. 

\begin{figure}[]
\centering

\begin{subfigure}{0.45\textwidth}
    \includegraphics[width=\textwidth]{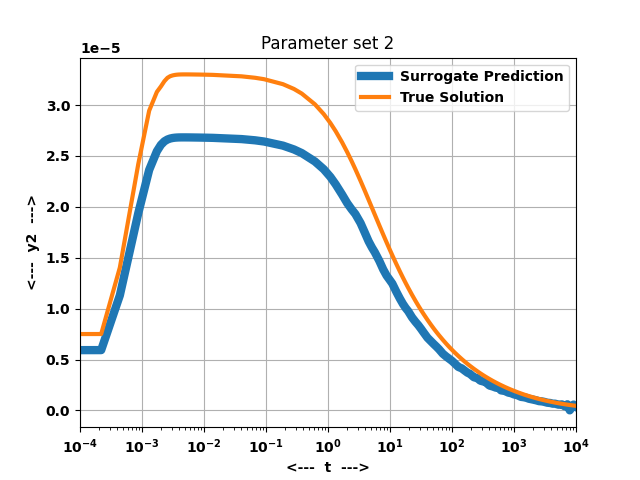}
    \caption{}
    \label{fig:nx_50_nd_50_no_poly_rate }
\end{subfigure}
\hfill
\begin{subfigure}{0.45\textwidth}
    \includegraphics[width=\textwidth]{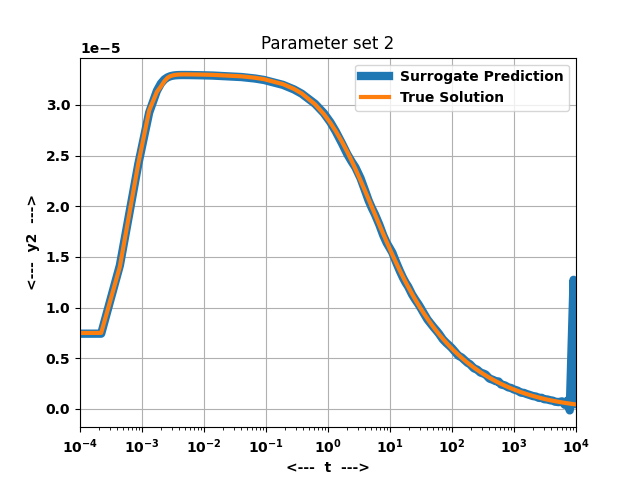}
    \caption{ }
    \label{fig:nx_50_nd_50_L_no_neigh_rate }
\end{subfigure}
\hfill
\par\medskip
\begin{subfigure}{0.45\textwidth}
    \includegraphics[width=\textwidth]{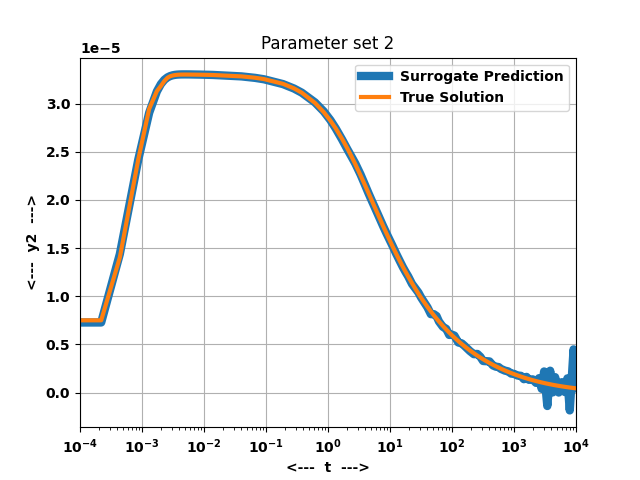}
    \caption{ }
    \label{fig:nx_50_nd_500_L_rate }
\end{subfigure}
\hfill
\begin{subfigure}{0.45\textwidth}
    \includegraphics[width=\textwidth]{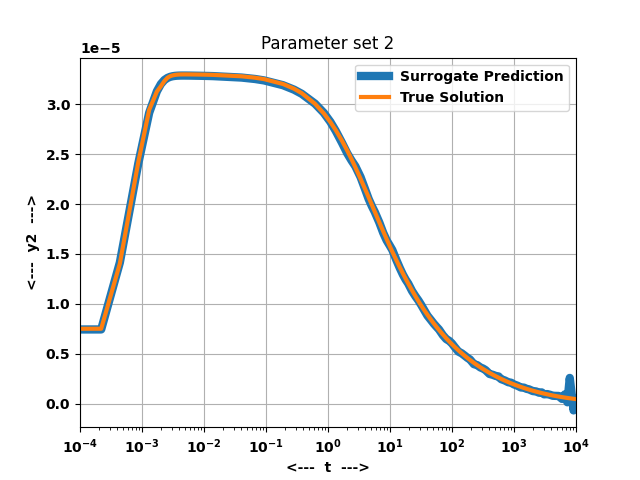}
    \caption{ }
    \label{fig:nx_50_nd_50_L_rate }
\end{subfigure}
\hfill
\par\medskip
\begin{subfigure}{0.45\textwidth}
    \includegraphics[width=\textwidth]{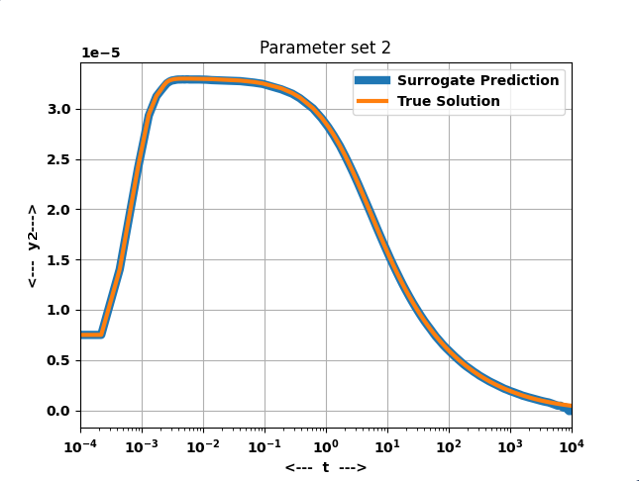}
    \caption{ }
    \label{fig:nx_50_nd_50_NL_rate }
\end{subfigure}
\hfill
\caption{\centering Figures show the time history for \(y_{2}\) for a test parameter set (P2 from Table \ref{table:rate_test_values}). Each trial mentioned below is referenced from Table \ref{table:robertson_rate_errors}. The NLPCTESN prediction is the best out of all models. (\subref{fig:nx_50_nd_50_no_poly_rate }) Trial 1 (\subref{fig:nx_50_nd_50_L_no_neigh_rate }) Trial 2
(\subref{fig:nx_50_nd_500_L_rate }) Trial 3
(\subref{fig:nx_50_nd_50_L_rate }) Trial 4
(\subref{fig:nx_50_nd_50_NL_rate }) Trial 5
}
\label{fig:rate_cond_plots}
\end{figure}

The average MAE for \(y_{2}\), averaged across several test cases listed in Table \ref{table:rate_test_values}, is shown in Table \ref{table:robertson_rate_errors} sorted in descending order of generalization MAE. Predictions for \(y_{2}\) for a particular test parameter are shown in Figure \ref{fig:rate_cond_plots}. For the same hyper-parameters, it can be observed that the absence of either the augmenting polynomial or the k-NN interpolation (i.e using all collocation points to predict the solution in the RBF) significantly increases the error of prediction. It can also be seen that the nonlinear projection performs better on average than the linear projection.

In the next section, the initial condition of Robertson's ODE is parametrized, and a similar error analysis is done.

\begin{table}[h]
\begin{center}
\begin{tabular}{ |c|c|c|c| } 
 \hline
 Parameter Set  & \(r_{1} / 0.04\)  & \(r_{2} / 3\cdot E7 \) & \(r_{3} / 1\cdot E4\)   \\ 
 \hline
  P1 & 0.95  & 1.05  & 0.95 \\ 
  \hline
  P2 & 0.9 & 1.1  & 0.9 \\ 
  \hline
  P3 & 1.1 & 0.9  & 1.1 \\ 
  \hline
  P4 & 1.03 & 0.99  & 1.04 \\ 
  \hline
\end{tabular}
\caption{\centering Test parameter values for rate parametrization of Robertson's ODE system}
\label{table:rate_test_values}
\end{center}
\end{table}

\begin{table}[h!]
\begin{center}
\begin{tabular}{ V{2} c|c|c|c|c|c|c V{2} } 
 \hlineB{2}
 \textbf{Trial No.} & \(\textbf{N}_r\) & \(\textbf{N}_{data}\)  & \(\textbf{N}_{neigh}\)  & \thead{\textbf{Polynomial}\\ \textbf{Augmentation}} & \thead{\textbf{Projection}\\ \textbf{Method}} & \textbf{Avg. MAE} (\(\times 10^{-7}\)) \\ 
 \hlineB{2}
  1 & 50 & 50  & 4 & N & Linear & 17.6 \\ 
  \hline
  2 & 50 & 50 & All & Y & Linear & 1.40\\
 \hline
  3 & 50 & 500  & 4 & Y & Linear & 1.29 \\
  \hline
  4 & 50 & 50 & 4 & Y & Linear & 0.76 \\
 \hline
  5 & 50 & 50 & 4 & Y & Nonlinear & \textbf{0.32}\\
  
 \hlineB{2}
\end{tabular}
\caption{\centering Average Mean Absolute Error (MAE) in \(y_{2}\), averaged across all test cases. }
\label{table:robertson_rate_errors}
\end{center}
\end{table}

\subsubsection{Parametrizing Initial Conditions}
\label{sec:rober_init}

The initial condition of the system is parameterized. This problem can be challenging for lower initial values of \(y_{1}\)(0) as it leads to a smaller and sharper transient in \(y_{2}\), which becomes more difficult to capture accurately by the surrogate. The DoE space of the initial condition is given as

\begin{equation}
\begin{aligned}
    \begin{split}
    y_{1}(0)&=[0.5,1],\\
    y_{2}(0)&=0,\\
    y_{3}(0)&=1-y_{1}(0).\\
    \end{split}
\end{aligned}
\end{equation}

The condition for \(y_{3}\) is decided on the basis that the sum of all quantities should always equal 1. Once more, the focus is on comparing the predicted results in \(y_{2}\).

\begin{table}[h]
\begin{center}
\begin{tabular}{ |c|c|c|c| } 
 \hline
 Parameter Set  & \(y_{1} \)  & \(y_{2} \) & \(y_{3}\)   \\ 
 \hline
  P1 & 0.6  & 0  & 0.4 \\ 
  \hline
  P2 & 0.7 & 0  & 0.3 \\ 
  \hline
  P3 & 0.85 & 0  & 0.15 \\ 
  \hline
  P4 & 0.9 & 0  & 0.1 \\ 
  \hline
\end{tabular}
\caption{\centering Test parameter values for initial condition parametrization of Robertson's ODE system}
\label{table:init_test_values}
\end{center}
\end{table}

\begin{table}[h!]
\begin{center}
\begin{tabular}{ V{2} c|c|c|c|c|c|c V{2} } 
 \hlineB{2}
 \textbf{Trial No.} & \(\textbf{N}_{r}\) & \(\textbf{N}_{data}\)  & \(\textbf{N}_{neigh}\)  & \thead{\textbf{Projection}\\ \textbf{Method}} & \textbf{Avg. MAE} (\(\times 10^{-7}\)) \\ 
 \hlineB{2}
  1 & 500 & 50  & All  & Linear & 158 \\ 
  \hline
  2 & 50 & 50 & All & Linear & 28.6\\
 \hline
  3 & 500 & 50 & 4  & Linear & 25.2 \\
 \hline
  4 & 50 & 50 & 4 & Linear & 6.23\\
 \hline
 5 & 50 & 50 & 4 & Nonlinear & \textbf{0.99}\\
 \hlineB{2}
\end{tabular}
\caption{\centering Average Mean Absolute Error (MAE) in \(y_{2}\), averaged across all test cases when parametrizing the initial condition. }
\label{table:robertson_init_errors}
\end{center}
\end{table}

\begin{figure}[]
\centering
\begin{subfigure}{0.45\textwidth}
    \includegraphics[width=\textwidth]{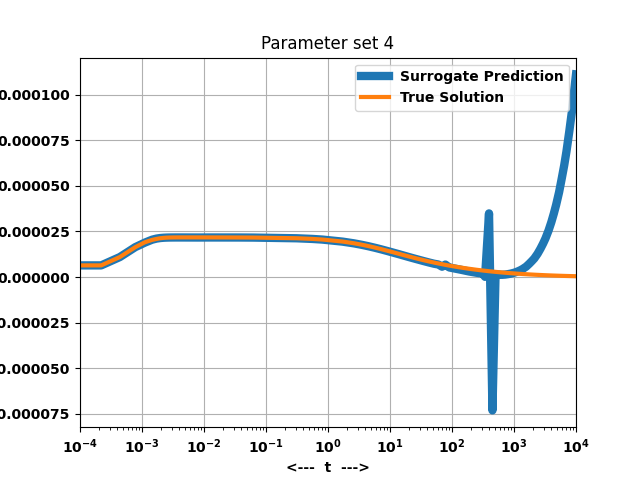}
    \caption{}
    \label{fig:nx_500_nd_50_L_no_neigh }
\end{subfigure}
\hfill
\begin{subfigure}{0.45\textwidth}
    \includegraphics[width=\textwidth]{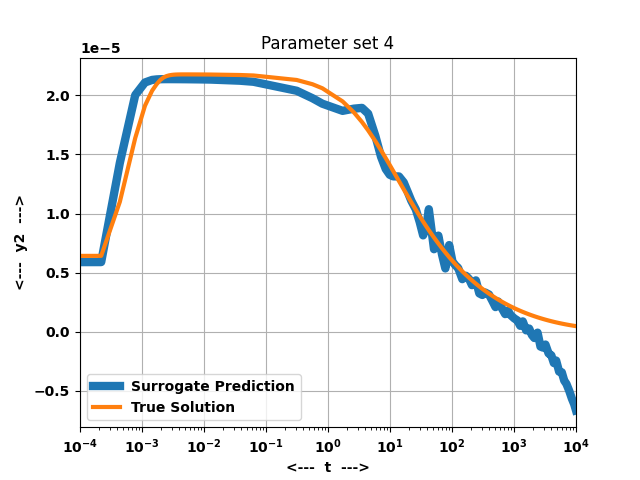}
    \caption{ }
    \label{fig:nx_50_nd_50_L_no_neigh }
\end{subfigure}
\hfill
\par\medskip
\begin{subfigure}{0.45\textwidth}
    \includegraphics[width=\textwidth]{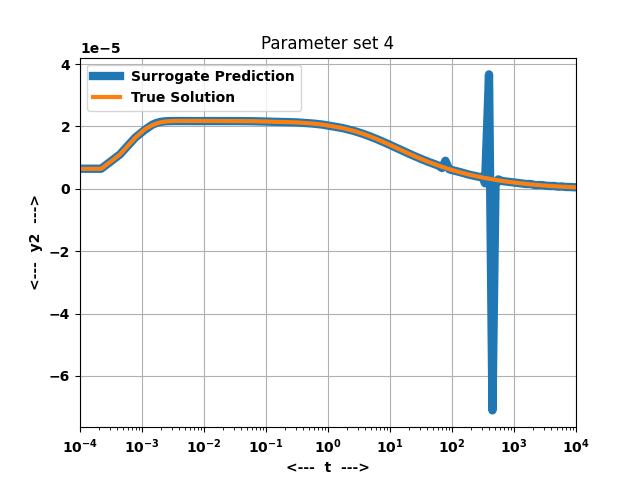}
    \caption{ }
    \label{fig:nx_500_nd_50_L }
\end{subfigure}
\hfill
\begin{subfigure}{0.45\textwidth}
    \includegraphics[width=\textwidth]{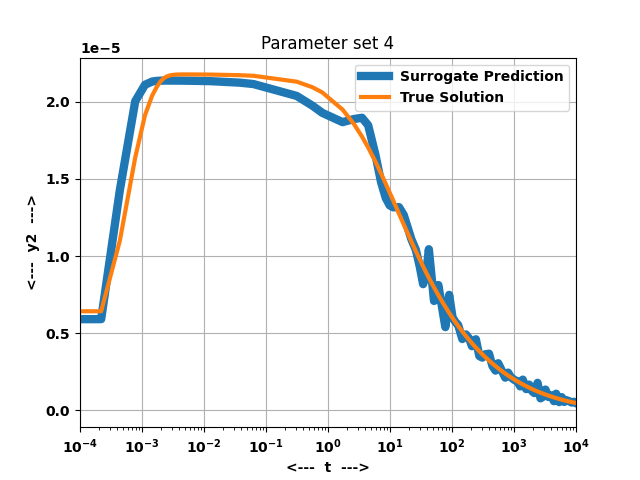}
    \caption{ }
    \label{fig:nx_50_nd_50_L }
\end{subfigure}
\hfill
\par\medskip
\begin{subfigure}{0.45\textwidth}
    \includegraphics[width=\textwidth]{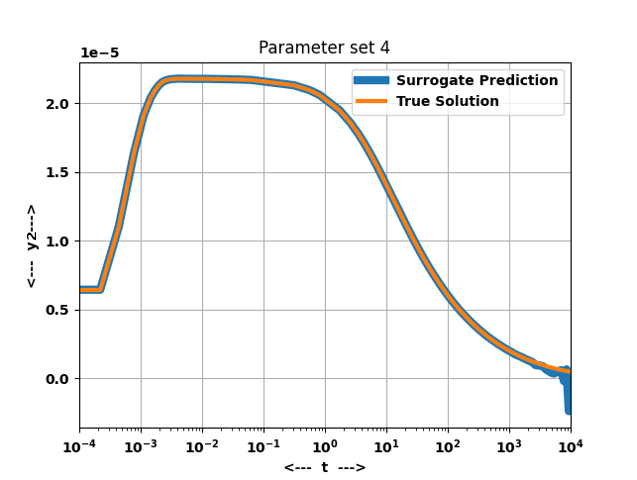}
    \caption{ }
    \label{fig:nx_50_nd_50_NL }
\end{subfigure}
\hfill
\caption{\centering Figures show time history for \(y_{2}\) for a test parameter set (P4 from Table \ref{table:init_test_values}). Each trial mentioned below is referenced from Table \ref{table:robertson_init_errors}. The NLPCTESN performs the best out of all models.  
(\subref{fig:nx_500_nd_50_L_no_neigh }) Trial 1 
(\subref{fig:nx_50_nd_50_L_no_neigh }) Trial 2
(\subref{fig:nx_500_nd_50_L }) Trial 3
(\subref{fig:nx_50_nd_50_L }) Trial 4
(\subref{fig:nx_50_nd_50_NL }) Trial 5
 }
\label{fig:init_cond_plots}
\end{figure} 

Table \ref{table:robertson_init_errors} tabulates the average MAE across all test cases for the problem, listed in Table \ref{table:init_test_values}, sorted in descending order. Figure \ref{fig:init_cond_plots} shows the comparison for a test parameter, across 5 different configurations of the CTESN surrogate model. A similar trend of hyper-parameter performance as seen in Table \ref{table:robertson_rate_errors} is noted, in that the absence of the k-NN interpolation increases the error of the prediction. For the same hyper-parameters, the nonlinear projection once again achieves lower generalization MAE than the linear projection.

\subsection{Sliding Basepoint Model}
\label{sec:sliding_basepoint}

\begin{figure}[h]
    \centering
    \includegraphics[angle=0,width=7 cm]{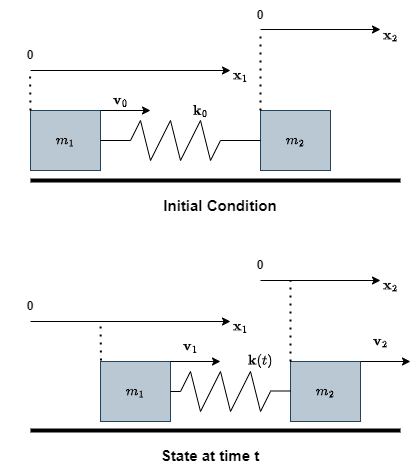}
    \caption{\centering Sliding Basepoint system for collision modeling.}
    \label{fig:crash_system_fig}
    \centering
    
\end{figure}

\begin{figure}[h!]
\centering
\begin{subfigure}{0.45\textwidth}
    \includegraphics[width=\textwidth]{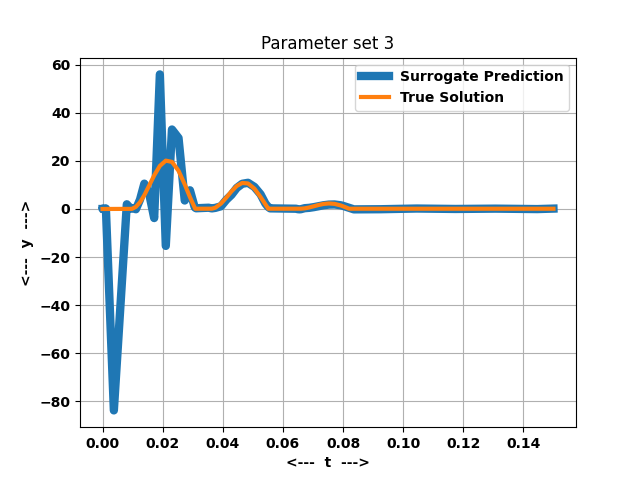}
    \caption{ }
    \label{fig:v2_2_LPCTESN }
\end{subfigure}
\hfill
\begin{subfigure}{0.45\textwidth}
    \includegraphics[width=\textwidth]{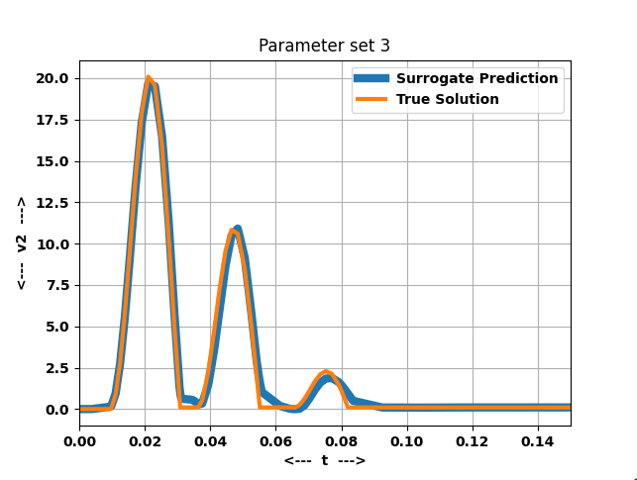}
    \caption{ }
    \label{fig:v2_2_NLPCTESN }
\end{subfigure}
\hfill
\par\medskip
\begin{subfigure}{0.45\textwidth}
    \includegraphics[width=\textwidth]{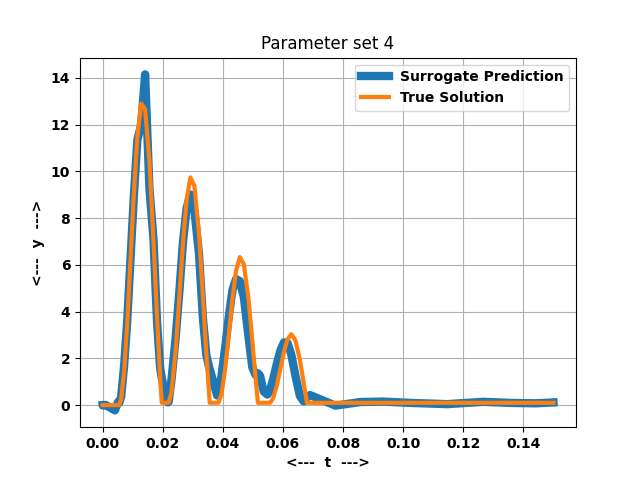}
    \caption{ }
    \label{fig:v2_3_LPCTESN }
\end{subfigure}
\hfill
\begin{subfigure}{0.45\textwidth}
    \includegraphics[width=\textwidth]{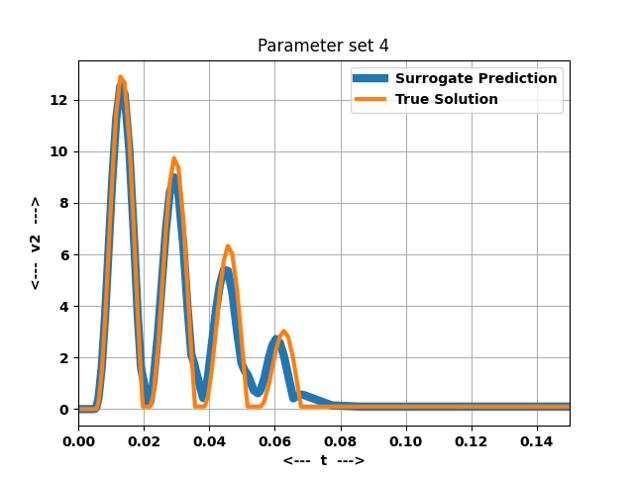}
    \caption{ }
    \label{fig:v2_3_NLPCTESN }
\end{subfigure}
\hfill
\par\medskip
\begin{subfigure}{0.45\textwidth}
    \includegraphics[width=\textwidth]{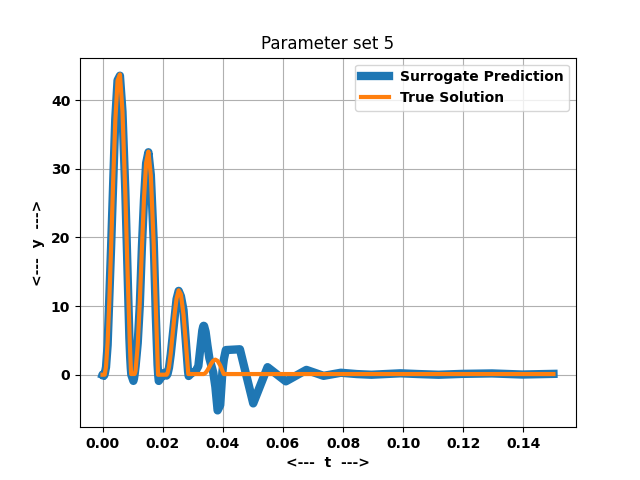}
    \caption{ }
    \label{fig:v2_4_LPCTESN }
\end{subfigure}
\hfill
\begin{subfigure}{0.45\textwidth}
    \includegraphics[width=\textwidth]{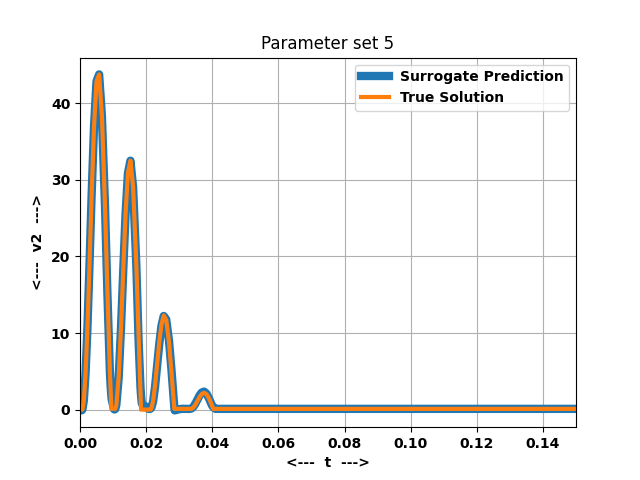}
    \caption{ }
    \label{fig:v2_4_NLPCTESN }
\end{subfigure}
\hfill

\caption{\centering Solution \(v_{2}\) for several tests with different test parameters, the average error of which is shown in Table \ref{table:crash_errors}. The left and right columns show results from the linear and nonlinear projections respectively. Table \ref{table: crash_test_values} defines the values P3-P5 mentioned below. (\subref{fig:v2_2_LPCTESN }) LPCTESN - P3 
(\subref{fig:v2_2_NLPCTESN }) NLPCTESN - P3 
(\subref{fig:v2_3_LPCTESN }) LPCTESN - P4  
(\subref{fig:v2_3_NLPCTESN }) NLPCTESN - P4 
(\subref{fig:v2_4_LPCTESN }) LPCTESN - P5  
(\subref{fig:v2_4_NLPCTESN }) NLPCTESN - P5 }
\label{fig:crash_plots_v2}
\end{figure}

Finally, the discussed methods are applied to the surrogate modeling of a realistic crash safety design problem. A system of ODEs proposed by \citet{horvath2012searching} that approximately models an automobile collision is solved via a created surrogate model. This system, called the Sliding Basepoint model, has parameters that were fitted on realistic crash data and are assumed to accurately model a collision problem. Figure \ref{fig:crash_system_fig} depicts the system at its initial state and at a later time. The system is given as:

\begin{equation}
    \begin{aligned}
    \begin{split}
    \dot{x_{1}}&=v_{1},\\
    \dot{v_{1}}&=\frac{1}{m_{1}}F_{1}(F_{s},c_{1},v_{1}),\\
    \dot{x_{2}}&=v_{2},\\
    \dot{v_{2}}&=\frac{1}{m_{2}}F_{2}(-F_{s},c_{2},v_{2}),\\
    \dot{k}&=D_{k}\cdot P,\\
    \dot{c_{1}}&=D_{c}\cdot P,\\
    \end{split}
    \end{aligned}
\end{equation}

where \(m_{1},m_{2},F_{1},F_{2},x_{1},x_{2},v_{1},v_{2}\) refer to the masses, total forces on, positions and velocities of the bodies respectively. \(m_{1}\) in reality reflects the mass of the chassis of the car, \(x_{1}\) behaves similar to the deformation of the bumper. \(F_{s}\) represents the spring force:

\begin{equation}
    F_{s}=k\cdot (x_{2}-x_{1}),
\end{equation}

between the two masses and P represents the power of dissipation

\begin{equation}
    P=|m_{1}\cdot v_{1}^{2}|.
\end{equation}

Finally, the forces are computed based on best-fit models described in the paper:

\begin{equation}
    \begin{aligned}
    \begin{split}
    F_{1}(F_{s},c_{1},v_{1})&=F_{s}-c_{1}\cdot |v_{1}|\cdot sign(v_{1}),\\
    F_{2}(F_{s},c_{2},v_{2})&=
    \begin{cases}
    0,              & \text{if } F_{s} < c_{2} \text{ or } |v_{2}|<v_{f}\\
    F_{s}-c_{2}\cdot sign(v_{2}),& \text{ otherwise }\\
    \end{cases}
    \end{split}
    \end{aligned}
\end{equation}

\begin{table}[h]
\begin{center}
\begin{tabular}{ |c|c|c|c| } 
 \hline
 Parameter Set  & \(v_{0}/25 \)  & \(k_{0}/ 10^{6}  \)   \\ 
 \hline
  P1 & 0.496   & 4.46 \\ 
  \hline
  P2 & 0.884  & 8.45 \\ 
  \hline
  P3 & 0.424  & 1.43 \\ 
  \hline
  P4 & 0.283  & 3.75 \\ 
  \hline
  P5 & 0.9  & 9.95 \\ 
  \hline
\end{tabular}
\caption{\centering Test parameter values for the collision modeling problem}
\label{table: crash_test_values}
\end{center}
\end{table}

\begin{table}[h]
\begin{center}
\begin{tabular}{ |c|c|c|c|c|c| } 
 \hlineB{2}
 \(\textbf{m}_{1}\) (kg) & \(\textbf{m}_{2}\) (kg)  & \(\textbf{D}_{k}\) & \(\textbf{D}_{c}\) & \(\textbf{c}_{2}\) (N) & \(\textbf{v}_{f}\) (m/s) \\ 
 \hlineB{2}
  1916 & 24.9 &  3.38 & 0.459 & 145000 & 0.1 \\ 
  \hline
\end{tabular}
\caption{\centering Parameter values taken from \cite{horvath2012searching} with their SI units.}
\label{table:crash_param_values}
\end{center}
\end{table}

The values of the fixed parameters are given in Table \ref{table:crash_param_values}. The reader is referred to the paper \cite{horvath2012searching} for further details. The system is subject to initial conditions \([x_{1},v_{1},x_{2},v_{2},k,c_{1}]=[0,v_{0},0,0,k_{0},0]\).

 The mechanics of crash and impact problems are known to have sharp transients and highly oscillatory behaviors associated with them, making their numerical solutions slow and costly to compute for a wide range of parameters. Hence, CTESNs will be a good surrogate modeling tool for the problem.

In this work, the initial conditions of the problem (\(v_{0}\) and \(k_{0}\)) are parametrized to simulate different impact velocities and directions (the spring constant of the bumper can be assumed to be different in different directions). The state space for the problem is the vector \([x_{1},v_{1},x_{2},v_{2},k,c_{1}]\).

 The DoE space is the range:

\begin{equation}
    \begin{aligned}
    \begin{split}
    \text{v}_{0} &= [5,25] \text{ (m/s)},\\ 
    \text{k}_{0}&=[1,10] \times 10^{6} \text{ kg}/s^{2},
    \end{split}
    \end{aligned}
\end{equation}

 chosen to represent a wide range of speeds and stiffness constants. The DoE space was sampled using a space-filling sampling strategy \cite{morris_mitchell} and the surrogate models were trained on 100 data points, and data was generated using a stiff ODE solver using solver parameters given in \cite{horvath2012searching}.
 
 Table \ref{table:crash_errors} shows the average MAE for the state variable \(v_{2}\) for the linear and nonlinear projection methods, all other hyper-parameters being kept the same, for test parameters listed in Table \ref{table:crash_param_values}. \(v_{2}\) was chosen to demonstrate the accuracy of the surrogates because, as visible in Figure \ref{fig:crash_plots_v2}, it has sharp nonlinear oscillatory transients starting from the moment of collision, which is difficult to capture for surrogate models. It can be seen from Table \ref{table:crash_errors} and Figure \ref{fig:crash_plots_v2} that the LPCTESN has a poorer performance on test cases compared to the NLPCTESN. 

Of particular interest is the speedup obtained by using the surrogate; using the surrogate leads to up to a 200x speedup in the prediction of the solution, the ODE solver taking roughly 0.02 seconds per solve of the ODE system. This is important because the transients in Figure \ref{fig:crash_plots_v2} occur at the same time scale \(\mathcal{O}(10^{-2}s)\). With a 200x speedup, if the surrogate is deployed on board a vehicle, it can judge the severity of the collision much quicker by computing impact forces from the result of the system, and closed-loop control and safety measures can be deployed. In this case, it would effectively function as a digital twin for collision monitoring.

\begin{table}[h!]
\begin{center}
\begin{tabular}{ V{2} c|c|c|c|c|c V{2} } 
 \hlineB{2}
 Trial No. & \(\textbf{N}_{x}\)  & \(\textbf{N}_{neigh}\)  & Projection Method  & Avg. MAE (\(v_{2}\)) \\ 
 \hlineB{2}
  1 & 50   & 4  & Linear & 1.02 \\ 
 \hline
  2 & 50  & 4  & Nonlinear & 0.51 \\
 \hlineB{2}
\end{tabular}
\caption{\centering Average Mean Absolute Error (MAE) in \(v_{2}\), averaged across all test cases in Table \ref{table: crash_test_values}}
\label{table:crash_errors}
\end{center}
\end{table}

\subsection{The POLLU Model}
\label{sec:POLLU}

\begin{table}[h]
\begin{center}
\begin{tabular}{ |L|L|L| }
\hline
\text{Species (y)}  & \text{Production Rate (P)} & \text{Loss Rate (L)} \\
\text{1}  & \thead{r_{2}y_{2}y_{4}+ r_{3}y_{5}y_{2} + r_{9}y_{11}y_{2} + r_{11}y_{13} \\ + r_{12}y_{10}y_{2} + r_{22}y_{19} + r_{25}y_{20}}  & \thead{r_{1}+r_{10}y_{11}+r_{14}y_{6} \\ +r_{23}y_{4}+r_{24}y_{19}} \\
\text{2}  & \thead{ r_{1}y_{1}+ r_{21}y_{19} }& \thead{r_{2}y_{4}+r_{3}y_{5}+r_{9}y_{11}+r_{12}y_{10}} \\
\text{3}  & \thead{ r_{1}y_{1}+ r_{17}y_{5}+ r_{19}y_{16}+ r_{22}y_{19} }& \thead{r_{15}} \\
\text{4}  & \thead{ r_{15}y_{3}}& \thead{r_{2}y_{2}+r_{16}+r_{17}+r_{23}y_{1}} \\
\text{5}  &\thead{ r_{4}y_{7}+r_{4}y_{7} +r_{6}y_{7}y_{6}+ r_{7}y_{9} \\ +r_{13}y_{14}+ r_{20}y_{17}y_{6}} & \thead{r_{3}y_{2}} \\
\text{6}  & \thead{ r_{3}y_{5}y_{2}+ r_{18}y_{16}+ r_{18}y_{16} }& \thead{r_{6}y_{7}+r_{8}y_{9}+r_{14}y_{1}+r_{20}y_{17}} \\
\text{7}  & \thead{ r_{13}y_{14}} & \thead{r_{4}+r_{5}+r_{6}y_{6}} \\
\text{8}  & \thead{ r_{4}y_{7}+ r_{5}y_{7}+ r_{6}y_{7}y_{6}+ r_{7}y_{9}} & \thead{0.0} \\
\text{9}  & \thead{ 0} & \thead{r_{7}+r_{8}y_{6}} \\
\text{10}  & \thead{ r_{7}y_{9}+ r_{9}y_{11}y_{2}} & \thead{r_{12}y_{2}} \\
\text{11}  & \thead{ r_{8}y_{9}y_{6}+ r_{11}y_{13}} & \thead{r_{9}y_{2}+r_{10}y_{1}} \\
\text{12}  & \thead{ r_{9}y_{11}y_{2}} & \thead{0.0} \\
\text{13}  & \thead{ r_{10}y_{11}y_{1}} & \thead{r_{11}} \\
\text{14}  & \thead{ r_{12}y_{10}y_{2}} & \thead{r_{13}} \\
\text{15}  & \thead{ r_{14}y_{1}y_{6}} & \thead{0.0} \\
\text{16}  & \thead{ r_{16}y_{4}} & \thead{r_{18}+r_{19}} \\
\text{17}  & \thead{ 0.0} & \thead{r_{20}y_{6}} \\
\text{18}  & \thead{ r_{20}y_{17}y_{6}} & \thead{0.0} \\
\text{19}  & \thead{ r_{23}y_{1}y_{4}+ r_{25}y_{20}} & \thead{r_{21}+r_{22}+r_{24}y_{1}} \\
\text{20} & \thead{ r_{24}y_{19}y_{1}} & \thead{r_{25}} \\
\hline
\end{tabular}
\caption{\centering Species involved in the POLLU reaction system, with their production and loss rates, derived from \cite{verwer1994gauss}.}
\label{table:POLLU_species_reactions}
\end{center}
\end{table}

The CTESN approach is used to model the POLLU air pollution model developed at the Dutch National Institute of Public Health and Environmental Protection \cite{verwer1994gauss}. It consists of 20 species and 25 reactions, modeled by nonlinear ODEs of the form

\begin{equation}
    \label{eqn:POLLU_model}
    \frac{dy}{dt}=P(t,y)-L(t,y)y
\end{equation}

where y is the concentration vector (in ppm) of the reacting species, P is the production term and L is a diagonal matrix representing the loss term for every species in the system. Table \ref{table:POLLU_species_reactions} shows the production and loss rates for each species of the system. The values of the rate constants \(r\) are given in table \ref{table:POLLU_rates}. The reader is referred to the paper \cite{verwer1994gauss} for a complete description of the reaction system.  

The system is a common benchmark for stiff ODE solvers and represents a difficult problem to solve due to the large number of species and ODEs involved. When such systems have to be solved at many grid points, say in a computational mesh, they represent a very expensive computation and hence this example is an ideal application for testing surrogates of stiff ODEs.
In this work, the initial conditions of the system are simultaneously parametrized, according to the following:

\begin{equation}
    \label{eqn:POLLU_doe}
    \begin{aligned}
    \begin{split}
    \text{y}_{2,0} &=  [0.14,0.26]  ,\\ 
    \text{y}_{4,0} &=  [0.028,0.052] ,\\
    \text{y}_{7,0} &=  [0.07,0.13] .\\
    \end{split}
    \end{aligned}
\end{equation}

where \(y_{2,0}\), \(y_{4,0}\) and \(y_{7,0}\) refer to the initial concentration of the respective species. The initial conditions for the rest of the species are default as per the paper \cite{verwer1994gauss}.

\begin{table}[h]
\begin{center}
\begin{tabular}{ |c|c|c|c| } 
 \hline
 Parameter Set  & \(y_{2,0}\)  & \(y_{4,0}\) & \(y_{7,0}\)   \\ 
 \hline
  P1 & 0.192   & 0.0464 & 0.114 \\ 
  \hline
  P2 & 0.258  & 0.04012 & 0.103 \\ 
  \hline
  P3 & 0.188  & 0.0332 & 0.0938 \\ 
  \hline
  P4 & 0.208  & 0.0508 & 0.115\\ 
  \hline
  P5 & 0.228  & 0.0488 & 0.0714\\ 
  \hline
\end{tabular}
\caption{\centering Test parameter values for the POLLU air pollution modeling problem}
\label{table:POLLU_test_params}
\end{center}
\end{table}

\begin{figure}[h!]
\centering
\begin{subfigure}{0.45\textwidth}
    \includegraphics[width=\textwidth]{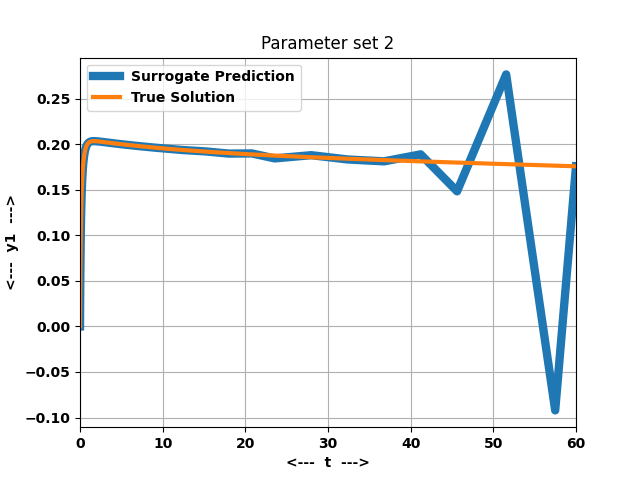}
    \caption{ }
    \label{fig:y1_1_LPCTESN}
\end{subfigure}
\hfill
\begin{subfigure}{0.45\textwidth}
    \includegraphics[width=\textwidth]{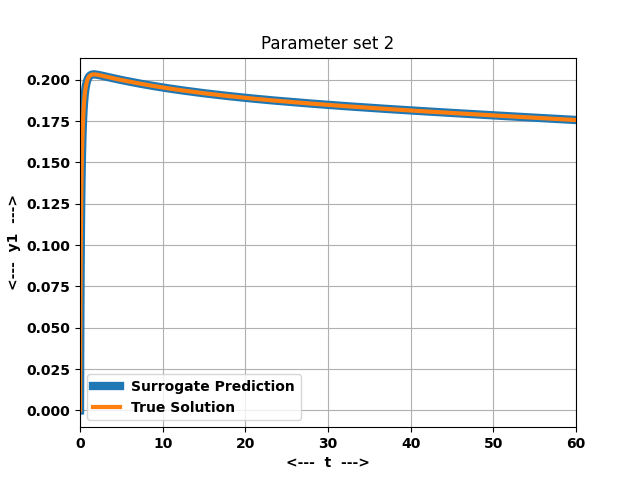}
    \caption{ }
    \label{fig:y1_1_NLPCTESN}
\end{subfigure}
\hfill
\par\medskip
\begin{subfigure}{0.45\textwidth}
    \includegraphics[width=\textwidth]{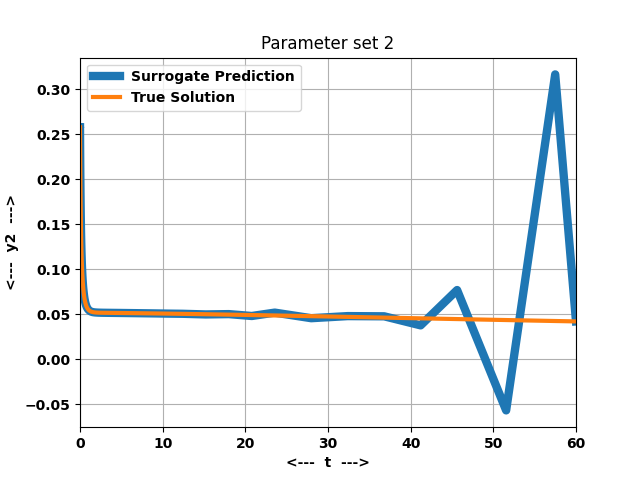}
    \caption{ }
    \label{fig:y2_1_LPCTESN}
\end{subfigure}
\hfill
\begin{subfigure}{0.45\textwidth}
    \includegraphics[width=\textwidth]{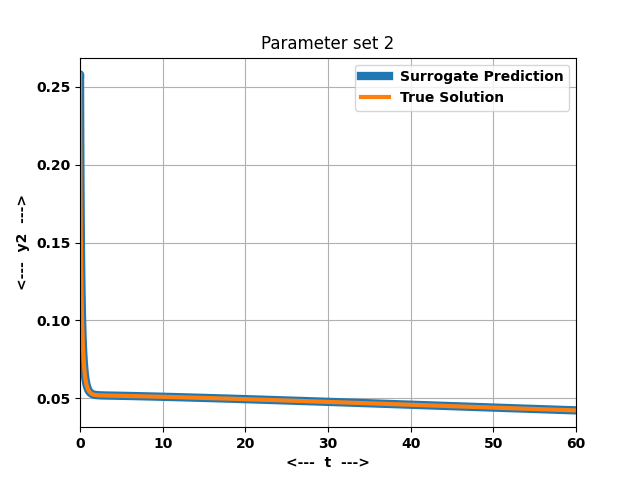}
    \caption{ }
    \label{fig:y2_1_NLPCTESN}
\end{subfigure}
\hfill
\par\medskip
\begin{subfigure}{0.45\textwidth}
    \includegraphics[width=\textwidth]{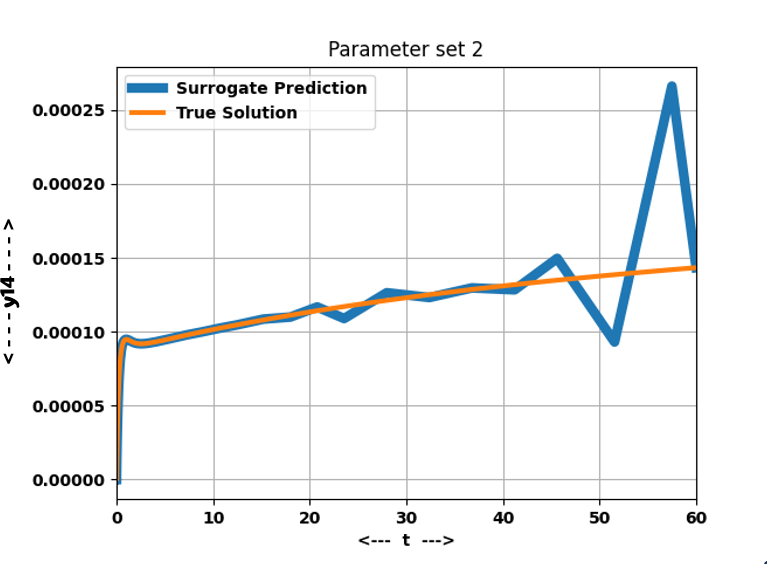}
    \caption{ }
    \label{fig:y14_1_LPCTESN}
\end{subfigure}
\hfill
\begin{subfigure}{0.45\textwidth}
    \includegraphics[width=\textwidth]{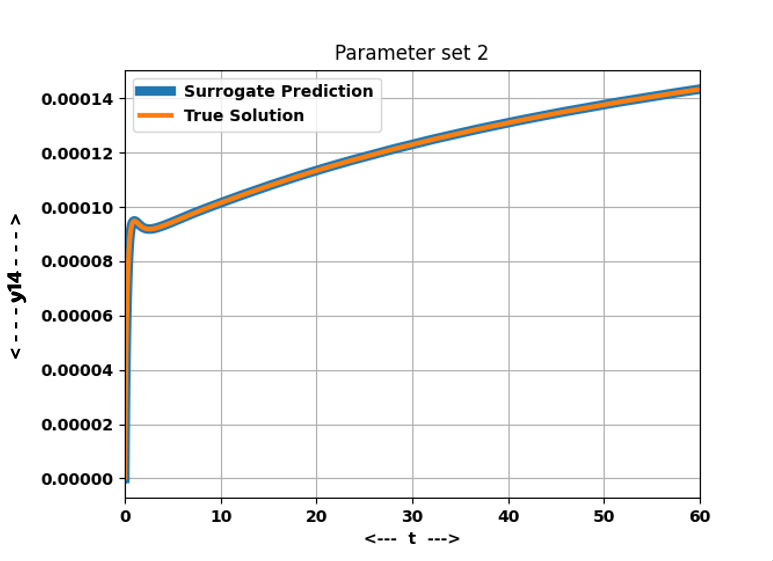}
    \caption{ }
    \label{fig:y14_1_NLPCTESN}
\end{subfigure}
\hfill
\par\medskip
\begin{subfigure}{0.45\textwidth}
    \includegraphics[width=\textwidth]{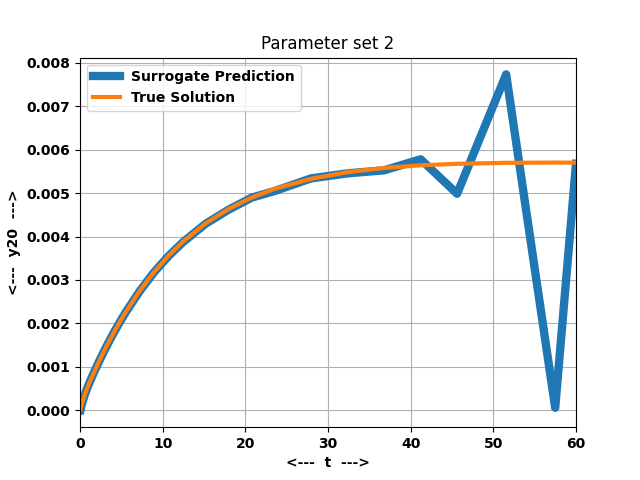}
    \caption{ }
    \label{fig:y20_1_LPCTESN}
\end{subfigure}
\hfill
\begin{subfigure}{0.45\textwidth}
    \includegraphics[width=\textwidth]{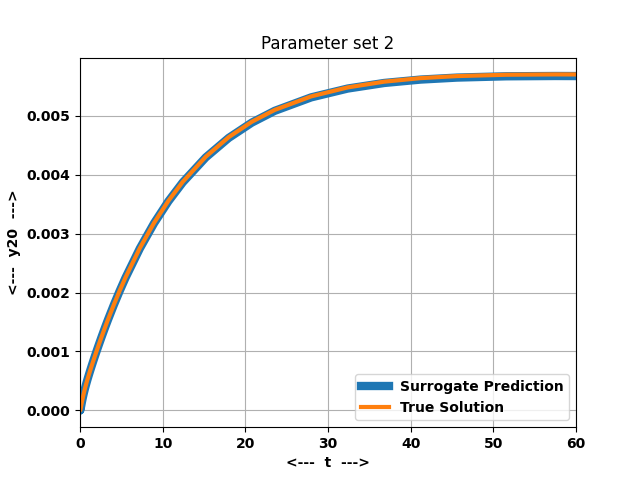}
    \caption{ }
    \label{fig:y20_1_NLPCTESN}
\end{subfigure}
\hfill

\caption{\centering Solutions \(y_{1}\),\(y_{2}\),\(y_{14}\) and \(y_{20}\) for a test parameter (P2 from Table \ref{table:POLLU_test_params}). The left and right columns contain results from the linear and nonlinear projections respectively. The NLPCTESN outperforms the LPCTESN in this example.(\subref{fig:y1_1_LPCTESN}) LPCTESN - \(y_{1}\) 
(\subref{fig:y1_1_NLPCTESN}) NLPCTESN - \(y_{1}\)
(\subref{fig:y2_1_LPCTESN}) LPCTESN - \(y_{2}\)
(\subref{fig:y2_1_NLPCTESN}) NLPCTESN - \(y_{2}\)
(\subref{fig:y14_1_LPCTESN}) LPCTESN - \(y_{14}\) 
(\subref{fig:y14_1_NLPCTESN}) NLPCTESN - \(y_{14}\)
(\subref{fig:y20_1_LPCTESN}) LPCTESN - \(y_{20}\)
(\subref{fig:y20_1_NLPCTESN}) NLPCTESN - \(y_{20}\) }
\label{fig:POLLU_plots}
\end{figure}

The training data was sampled using 100 data points within this DoE space, selected randomly. The reservoir size \(\textbf{N}_{r}\) was set to 100, and the number of queried neighbors \(\textbf{N}_{neigh}\) by the k-NN RBF was set to 10. Table \ref{table:POLLU_errors} shows the mean absolute errors computed across several test cases (listed in Table \ref{table:POLLU_test_params}) for a few species in the reaction, for the linear and nonlinear projection methods. It can be observed that the nonlinear projection CTESN outperforms the linear projection CTESN when all other hyper-parameters are kept the same. Figure \ref{fig:POLLU_plots} shows the comparison of the prediction graphically; the LPCTESN prediction is much noisier than the NLPCTESN prediction at later times. This was also observed with Robertson's equations, where the predictions at larger time scales by the LPCTESN tended to become noisier.

\begin{table}[h!]
\begin{center}
\def\arraystretch{1.2}
\begin{tabular}{ V{2} c|c|c|c|c|c V{2} } 
 \hlineB{2}
  \multirow{2}{*}{\thead{\textbf{Projection } \\ \textbf{Method}}}  & \multicolumn{5}{c V{2}}{\textbf{Mean Absolute Error}} \\\cline{2-6} 
 
     & \(y_{1} (\times 10^{-5})\) & \(y_{2} (\times 10^{-5})\) & \(y_{4} (\times 10^{-4})\) & \(y_{14} (\times 10^{-8})\) & \(y_{20} (\times 10^{-6})\)\\
 \hlineB{2}
    Linear & 75.1 & 76.3 & 7.68 & 51.5 & 22.1 \\ 
 \hline
    Nonlinear & 2.60 & 2.78 & 3.01 & 1.76 & 2.59 \\
 \hlineB{2}
\end{tabular}
\caption{\centering Mean Absolute Error (MAE) for the POLLU problem, averaged across all test cases in Table \ref{table:POLLU_test_params}.}
\label{table:POLLU_errors}
\end{center}
\end{table}

\section{Discussion and Conclusion}
\label{sec:conclusion}

From the numerical experiments conducted, it was observed that polynomial-augmented K-nearest neighbor RBFs outperform standard RBFs in terms of accuracy when used as part of the CTESN surrogate modeling algorithm. It was observed that the nonlinear projection (NLPCTESN) method consistently outperformed the linear projection (LPCTESN) method, achieving superior accuracy when the hyper-parameters are the same, on a variety of problems. The NLPCTESN method demonstrated accuracy across several problems with sharp transients, varying time scales, and long horizons of integration, whereas the LPCTESN method had a higher generalization error on all test examples. The surrogate was also shown to achieve a speedup of several orders of magnitude compared to an ODE solver of a realistic collision problem and could be used for several cases like design optimization and online collision severity monitoring.

There are several directions in which this work could be built upon. The CTESN is a black-box data-driven method, and future works need to investigate how well the model learns the physics of the problem, or apply physics-constrained modeling approaches to the outputs of the CTESN. The speedup of the surrogate model becomes very apparent when solving many instances of the ODE; this happens either when the ODE system is very large, or the small ODE system has to be solved repeatedly many times, such as in coupled ODE-PDE systems. Examples include chemically reacting flows similar to the POLLU system, in which a large stiff system of ODEs has to be solved at every compute node, or FEM-based crash solvers which require accurate modeling of sharp transients similar to those demonstrated in this work.\\

\bibliographystyle{unsrtnat}
\bibliography{bib}
\appendix

\section{k-Nearest Neighbor Polynomial Augmented Radial Basis Function Interpolation}
\label{sec:NN_RBF_Poly}

Radial basis function (RBF) interpolation is a high-order accurate method that uses radial basis functions to create interpolants of unstructured data, which can be in an arbitrary number of dimensions. The scalar form of the interpolant is

\begin{equation}
    s(x)=\sum_{i=0}^{N} w_{i} \phi(||x-x_{i}||_{2})
\end{equation}

where \(x_{i}\) represent the points at which the solution is known, \(w_{i}\) are the associated coefficients which are fitted and \(\phi\) is a kernel function. One of the most common kernel functions, also used in this work, is

\begin{equation}
    \phi(\epsilon r)= (\epsilon r)^{2}log(\epsilon r)
\end{equation}

where \(\epsilon\) is the shape factor, which is an important hyperparameter. In this work, it is set to 1. However, it can often greatly affect the generalization capability of the RBF. \citet{cao2022polynomial} discussed how high-order polynomial augmented RBFs outperformed standalone RBFs and reduced the dependency of the RBF on the shape factor \(\epsilon\), and several works \cite{jankowska2018improved,yao2017modified} have shown that adding high-order polynomials to the RBF greatly enhances the accuracy of the model. In this work, the polynomial augmented RBF takes the form:

\begin{equation}
    f(x)=\textbf{K}(\textbf{x},\textbf{y})\textbf{A}+\textbf{P}(\textbf{x})\textbf{B}
\end{equation}

where \textbf{A} and \textbf{B} are fitted coefficient matrices, and \textbf{K} and \textbf{P} need to be constructed for a query point x, given data points \textbf{y}. If \(N_{d}\), \(N_{O}\), \(N_{p}\) are determined by the number of data points, the output dimension, and the polynomial order respectively, we have \textbf{A} \(\in \mathbb{R}^{N_{d}\times N_{O}}\) and \textbf{B} \(\in \mathbb{R}^{N_{p}\times N_{O}}\). \textbf{K} \(\in \mathbb{R}^{N_{d}}\) can be constructed as

\begin{equation}
    \label{eqn:K}
    K_{i}(x)=\phi(||x-y_{i}||), \text{ i=1....}N_{d},
\end{equation}

and \textbf{P} \(\in \mathbb{R}^{N_{p}}\) can be constructed as

\begin{equation}
    \label{eqn:P}
    \textbf{P}(\textbf{x})= \mathbbm{1}+\textbf{x}+....
\end{equation}

A further improvement to RBF interpolation is adding a k-Nearest Neighbor constraint to the prediction process. This means that during prediction, only k of the nearest collocation points ( i.e \(y_{i}\)) will contribute to the prediction at the test point. These N points in practice are usually inferred using a decision tree. Intuitively, this method is useful when the interpolation spaces are large and the hyper-parameters of the RBF may not be optimally tuned, leading to collocation points at large distances from the test point and contributing to the interpolant evaluation, corrupting the solution.

\section{POLLU Reaction rates}

The rates of reaction of the POLLU system are shown in Table \ref{table:POLLU_rates}.

\begin{table}[h]
\begin{center}
\begin{tabular}{|L|L|L|L|}
\hline
\text{Reaction Rate} & \text{Value} & \text{Reaction Rate} & \text{Value} \\
 r_{1} & 0.350\text{E}+0 & r_{14} & 0.163\text{E}+5 \\
 r_{2} & 0.266\text{E}+2 &  r_{15} & 0.480\text{E}+7 \\
 r_{3} & 0.120\text{E}+5 & r_{16} & 0.350\text{E}-3 \\
 r_{4} & 0.860\text{E}-3 & r_{17} & 0.175\text{E}-1 \\
 r_{5} & 0.820\text{E}-3 & r_{18} & 0.100\text{E}+9 \\
 r_{6} & 0.150\text{E}+5 &  r_{19} & 0.444\text{E}+12 \\
 r_{7} & 0.130\text{E}-3 &  r_{20} & 0.124\text{E}+4 \\
 r_{8} & 0.240\text{E}+5 &  r_{21} & 0.210\text{E}+1 \\
 r_{9} & 0.165\text{E}+5 &   r_{22} & 0.578\text{E}+1 \\
 r_{10} & 0.900\text{E}+4 &   r_{23} & 0.474\text{E}-1 \\
 r_{11} & 0.220\text{E}-1 &   r_{24} & 0.178\text{E}+4 \\
 r_{12} & 0.120\text{E}+5 &  r_{25} & 0.312\text{E}+1 \\
 r_{13} & 0.188\text{E}+1 &          & \\

\hline
\end{tabular}
\caption{\centering Reaction rates for the POLLU problem}
\label{table:POLLU_rates}
\end{center}
\end{table}

\end{document}